\documentclass[12pt]{article}
\usepackage{jheppub}
\pdfoutput=1
\usepackage[utf8]{inputenc}
\usepackage{comment}
\usepackage[absolute]{textpos}
\input epsf.sty
\usepackage[cal=boondoxo]{mathalfa}
\usepackage{verbatim}         
\usepackage{graphicx}						
\usepackage{hyperref}
\usepackage{amssymb}
\usepackage{amsmath}
\usepackage{xcolor}
\hypersetup{
  colorlinks   = true, 
  urlcolor     = blue, 
  linkcolor    = blue, 
  citecolor   = red  
}
\usepackage{braket}
\usepackage{amssymb}
\usepackage{amsmath}
\usepackage{amsthm}
\usepackage{mathrsfs}
\usepackage{skak}
\usepackage{framed}
\usepackage{hyperref}
\usepackage{slashed}
\usepackage{array}

\hypersetup{colorlinks=true}
\hypersetup{linkcolor=blue}
\hypersetup{citecolor=blue}
\hypersetup{urlcolor=blue}

\numberwithin{equation}{section}

\usepackage
{datetime}
\usepackage{fancyhdr}
\usepackage{tikz, pgf}
\usetikzlibrary{shapes.misc}
\usetikzlibrary{shapes}
\usetikzlibrary{fit}
\usetikzlibrary{arrows}

\usetikzlibrary{decorations.pathmorphing}	
\usetikzlibrary{decorations.markings}
 \usetikzlibrary{patterns}

\tikzset{
    sugra/.style={decorate, decoration={snake}, draw=black},
    scalarphi/.style={dashed,draw=black, postaction={decorate},
        },
    scalarchi/.style={draw=brown}, 
    hwbou/.style={draw=blue, postaction={decorate}, ultra thick
        },
    vector/.style={draw=blue,decorate, decoration={snake}, draw},
	provector/.style={decorate, decoration={snake,amplitude=2.5pt}, draw},
	antivector/.style={decorate, decoration={snake,amplitude=-2.5pt}, draw},
   	 fermion/.style={draw=cyan, postaction={decorate},
        decoration={markings,mark=at position .55 with {\arrow[draw=black]{>}}}},
    fermionbar/.style={draw=cyan, postaction={decorate},
        decoration={markings,mark=at position .55 with {\arrow[draw=black]{<}}}},
    fermionnoarrow/.style={draw=black},
    gluon/.style={decorate, draw=red,
        decoration={coil, amplitude=4pt, segment length=5pt}},
    scalar/.style={dashed,draw=black, postaction={decorate},
        decoration={markings,mark=at position .55 with {\arrow[draw=black]{>}}}},
    scalarbar/.style={dashed,draw=black, postaction={decorate},
        decoration={markings,mark=at position .55 with {\arrow[draw=black]{<}}}},
    electron/.style={draw=black, postaction={decorate},
        decoration={markings,mark=at position .55 with {\arrow[draw=black]{>}}}},
    scalarnoarrow/.style={dashed, draw=black},
    electron/.style={draw=black, postaction={decorate},
        decoration={markings, mark=at position .55 with {\arrow[draw=black]{>}}}},
	bigvector/.style={decorate, decoration={snake, amplitude=4pt}, draw},
    photon/.style={draw=red, decorate, decoration={snake}, draw},
    higgs/.style={dashed, draw=black, postaction={decorate},
        },	
        goldstone/.style={draw=brown, postaction={decorate},
        },    
          ghost/.style={dashed, draw=magenta, postaction={decorate},
        decoration={markings, mark=at position .55 with {\arrow[draw=black]{>}}}
        },  
          antighost/.style={dashed, draw=magenta, postaction={decorate},
        decoration={markings, mark=at position .55 with {\arrow[draw=black]{<}}}
        }, 
            scalartwo/.style={dashed,draw=brown, postaction={decorate},
        decoration={markings,mark=at position .55 with {\arrow[draw=black]{>}}}},
    scalarbartwo/.style={dashed,draw=brown, postaction={decorate},
        decoration={markings,mark=at position .55 with {\arrow[draw=black]{<}}}}, 
    fermiontwo/.style={draw=purple, postaction={decorate},
        decoration={markings,mark=at position .55 with {\arrow[draw=black]{>}}}},
    fermionbartwo/.style={draw=purple, postaction={decorate},
        decoration={markings,mark=at position .55 with {\arrow[draw=black]{<}}}},    
        mphoton/.style={decorate, decoration={snake}, draw=violet},
        realscalar/.style={draw=black}, 
        fakerealscalar/.style={draw=white}, 
        realscalarone/.style={ draw=black},
    	realscalartwo/.style={draw=brown},    	    pseudoscalar/.style={draw=brown},
        mgluon/.style={decorate, draw=blue,
        	decoration={coil,amplitude=4pt, segment length=5pt}},
         weylfermion/.style={draw=orange, postaction={decorate},
        decoration={markings,mark=at position .55 with {\arrow[draw=black]{>}}}},
         weylfermionbar/.style={draw=orange, postaction={decorate},
        decoration={markings,mark=at position .55 with {\arrow[draw=black]{<}}}}, 
    majorana/.style={draw=cyan, postaction={decorate},
        decoration={markings,mark=at position .55 with {\arrow[draw=black]{><}}}},
    majoranabar/.style={draw=cyan, postaction={decorate},
        decoration={markings,mark=at position .55 with {\arrow[draw=black]{><}}}},    
   	wboson/.style={draw=blue,decorate, decoration={snake,amplitude=4pt}, draw},  
    zboson/.style={draw=violet, decorate, decoration={snake}, draw},   
    lepton/.style={draw=black, postaction={decorate},
        decoration={markings, mark=at position .55 with {\arrow[draw=black]{>}}}},
    leptonbar/.style={draw=black, postaction={decorate},
        decoration={markings, mark=at position .55 with {\arrow[draw=black]{<}}}}, 
    clepton/.style={draw=cyan, postaction={decorate},
        decoration={markings, mark=at position .55 with {\arrow[draw= black]{>}}}},
    cleptonbar/.style={draw=cyan, postaction={decorate},
        decoration={markings, mark=at position .55 with {\arrow[draw=black]{<}}}},   
   nlepton/.style={draw=orange, postaction={decorate},
        decoration={markings, mark=at position .55 with {\arrow[draw=black]{>}}}},
    nleptonbar/.style={draw=orange, postaction={decorate},
        decoration={markings, mark=at position .55 with {\arrow[draw=black]{<}}}},              
        graviton/.style={draw=blue, decorate, decoration={snake, amplitude=3pt}, draw},  
        bgraviton/.style={draw=blue, decorate, decoration={snake, amplitude=4pt},  draw},  
        gravitino/.style={draw=red, postaction={decorate}, 
        decoration={snake,  markings, mark=at position .55 with {\arrow[draw=black]{><}}}},
    	gravitinobar/.style={draw=red, postaction={decorate},
        decoration={snake, markings, mark=at position .55 with {\arrow[draw=black]{><}}} },  
    phir/.style={draw=blue, postaction={decorate},},
   phil/.style={dashed,draw=blue,},
     phiav/.style={draw=cyan, postaction={decorate},},
   phidif/.style={dashed,draw=cyan,},  
    chir/.style={draw=red, postaction={decorate},},
   chil/.style={dashed,draw=red,},  
}

\tikzstyle{block} = [draw, rectangle, 
    minimum height=3em, minimum width=6em]

\newcommand{\be}{\begin{eqnarray}\displaystyle}
\newcommand{\ee}{\end{eqnarray}}
\newcommand{\nn}{\nonumber}
\newcommand{\f}{\frac}
\newcommand{\p}{\partial}

\title{Perturbative soft photon theorems in de Sitter spacetime  }
\author[1]{Sayali Bhatkar,}
\emailAdd{sayali.bhatkar@tifr.res.in}

\author[1]{Diksha Jain}
\emailAdd{diksha.jain@tifr.res.in}

\affiliation[1]{Tata Institute of Fundamental Research\\
Dr Homi Bhabha Road, Navy Nagar, Mumbai, 400005, India
\vspace{4pt}}

\begin{abstract}
{We define a perturbative S-matrix in a local patch of de Sitter background in the limit when the curvature length scale ($\ell$) is large and study the 'soft' behavior of the scalar QED amplitudes in de Sitter spacetime in generic dimensions. We obtain the leading and subleading perturbative corrections to flat space soft photon theorems in the large $\ell$ limit, and comment on the universality of these corrections. We compare our results with the electromagnetic memory tails obtained earlier in $d=4$ using classical radiation analysis.}
\end{abstract}

\begin{document}

\maketitle

\newpage

\vspace{2cm}
\section{Introduction}
Gauge theories in flat spacetime display remarkable universal properties in the infrared regime. One manifestation of this universality are the so called soft theorems \cite{Bloch:1937pw,Gell-Mann:1954wra,Low:1954kd,Weinberg:1965nx,PhysRev.166.1287,PhysRev.168.1623,White:2011yy}. In a scattering process with $n$ hard (finite energy) particles and a soft ($k_\mu \rightarrow 0$) particle, the leading term in the soft expansion of scattering amplitudes goes like inverse of the soft energy and the coefficient of this term is a universal soft factor times the lower $n$-point amplitude without the soft particle. In the case when the soft particle is a photon, the soft factor depends only on the electric charge and momenta of the hard particles and is completely insensitive to other details of the process. It has been shown that the leading soft photon theorem is equivalent to Ward identity of asymptotic symmetry of QED \cite{FERRARI1971316,FERRARI1970553,He:2014cra,Campiglia:2015qka,Kapec:2015ena}. This line of study has been extended beyond the leading order as well \cite{Campiglia:2016hvg,Lysov:2014csa}.

Another cornerstone of infrared physics of gauge theories involves the memory effects \cite{Susskind:2015hpa, Bieri:2013hqa,Pasterski:2015zua}. These are classical observables defined at late times and are fixed in terms of the soft factors appearing in the soft limit of amplitudes. It has been shown in the context of QED that the leading soft photon theorem gives rise to a kick in the velocity of an asymptotic detector and is known as the electromagnetic memory effect. The general relation between classical soft radiation emitted in a scattering process and quantum soft factors appearing in the soft limit of scattering amplitudes was studied in \cite{Laddha:2018rle}.

Since we live in an expanding universe, a natural question is to study the effect of the cosmological constant ($\Lambda$) on the infrared physics of gauge theories. The early universe with constant scale factor can be well approximated by de Sitter spacetime, and there has been an extensive study of the 'soft' limit of in-in correlators defined in Poincare patch of de Sitter spacetime   \cite{Maldacena:2002vr,Creminelli:2012ed,Assassi:2012zq,Kundu:2014gxa,Ghosh:2014kba}\footnote{Interested readers can find a more exhaustive list of references in \cite{Armstrong:2022vgl}.}. These soft theorems relate the higher point in-in correlation functions to lower point correlation functions via the symmetries of the theory. Another natural quantity to consider in de Sitter spacetime is the wavefunction. Soft limits of wavefunction coefficients have been explored in \cite{Armstrong:2022vgl}. These notions of soft limits are not directly related to the soft limit of S-matrices in flat spacetime. In the present paper, we aim to analyze the soft limit of perturbative S-matrix in de Sitter spacetime that admits a straightforward flat space limit. Memory effects have also been explored in de Sitter spacetime in \cite{Bieri:2015jwa,Chu:2016qxp,Tolish:2016ggo,Hamada:2017gdg,dSmem5}. The relation between these memory effects and the soft limits of inflationary correlators is an open question.

In this paper, we study the soft photon theorem in de Sitter background by treating the cosmological constant as a perturbative parameter. Since the observable universe in de Sitter spacetime is confined to the cosmological horizons, we will study scattering processes inside the static patch \cite{Albrychiewicz:2020ruh}. Even the detectors lie within the same region, and hence we define the corresponding S-matrix in a small region \cite{S} inside the static patch.  It is assumed that all the length scales of our problem are much smaller than the de Sitter curvature length $\ell$ where $\ell$ is related to cosmological constant as $\Lambda=\f{3}{\ell^2}$. It should be emphasized that this perturbative definition of the S-matrix is different from the global S-matrix defined in de Sitter spacetime in \cite{Marolf:2012kh}. Defining global S-matrix in de Sitter spacetime is very non-trivial due to the absence of a single observer which can access the full spacetime \cite{Bousso:2004tv}. Several attempts have been made to precisely define the S-matrix in de Sitter spacetime. But our construction is less ambitious since we use a bottom-up approach to study perturbative corrections to flat space amplitudes.

\subsection{Main results}
We summarize the main results of our paper below.  We derived the leading and subleading corrections to flat space soft factor in generic spacetime dimension. Throughout this paper, by leading (subleading) order, we mean $O\left(\f{1}{\ell} \right)\left(O\left(\f{1}{\ell^2}\right)\right)$. This terminology should not be confused with the usual notion of leading and subleading order in soft momentum $k$. 

We study the S-matrix amplitude corresponding to the scattering of modes that reduce to plane waves in the limit $\ell\rightarrow\infty$. Particle production is exponentially suppressed in this limit and does not play a role in our analysis. As expected, we reproduce the flat space soft photon theorem at zeroth order in $\f{1}{\ell}$ expansion. The effect of de Sitter potential on the hard particles and the soft photon is captured at subsequent orders in $\f{1}{\ell}$ expansion. Unlike the flat space case the particles continue to accelerate at late times in de Sitter spacetime. This should be visible in the form of new non-analytic terms in the soft limit of scattering amplitudes. And indeed, we find that new soft modes appear at order $\f{1}{k^3\ell^2}$ in general spacetime dimension. Our results are listed below. 
\be
\Gamma_{n + 1}(\{p_i\}, k) = \left(S^{(0)} + \f{1}{\ell} S^{(1)} + \f{1}{\ell^2} S^{(2)} \right) \ \Gamma_{n}(\{p_i\}),\nn
\ee 
where $$
S^{(0)} =   \sum_{i = 1}^{n} e_i\frac{\varepsilon . p_i}{p_i.k} + O(k^0),
$$ 
\be
S^{(1)} =  -im \sum_{i = 1}^{n} e_i\left(\varepsilon .p_i\frac{k. \p_{p_i}}{ (k.p_i)^2}-  \frac{\varepsilon. \p_{p_i} }{ k.p_i} \right) + O(k^0)\nn
\ee
and
\be
\begin{split}\label{2s}
S^{(2)} = \sum_{i = 1}^{n}   e_i &\left(-m^2\frac{(d-4) }{4} \frac{\varepsilon .p_i}{(p_i.k)^3} + i m^2\frac{(d-4) }{4} \frac{\varepsilon .p_i}{(p_i.k)^3}k.\p_{p_i} + \f{m^2}{2} \left(\varepsilon .p_i\frac{k. \p_{p_i}}{ (p_i.k)^3}-  \frac{\varepsilon. \p_{p_i} }{ (p_i.k)^2}  \right)\right.\nn\\
&+ \frac{ (d-4)}{4}\frac{\varepsilon .p_i}{(p_i.k)^2}\left. - \frac{(d-4) }{4} \frac{\varepsilon .p_i}{(p_i.k)^2}p_i .\p_{p_i}\right) + O \left(\f{1}{k}\right).
\end{split}
\ee
where $\Gamma_{n + 1}(\{p_i\}, k)$ is the scattering amplitude of a process with $n$ hard particles of mass $m$ carrying momentum $p_i$, charge $e_i$ and a soft photon carrying momentum $k$ in $d$ dimensions. In our convention, $p_i$, $k$ and $e_i$ include an extra minus sign if the corresponding particle is incoming. $\varepsilon_\mu$ denotes the polarization vector of the external soft photon and its helicity index has been suppressed. The leading and subleading soft factors are denoted by $S^{(1)}$ and $S^{(2)}$ respectively and $S^{(0)}$ is the usual flat space soft factor.
\begin{itemize}
    \item As explained in \S \ref{5.22}, the subleading soft factor $S^{(2)}$ is universal and it is related to the memory tails obtained in \cite{2108} in $d = 4$. From \eqref{2s} notice that in $d=4$, the subleading soft factor starts at $\f{1}{k^2}$. This simplification occurs because Maxwell Lagrangian enjoys Weyl invariance in four dimensions. 
    \item On the other hand, the leading soft factor $S^{(1)}$ is not universal and is absent in classical analysis. We elaborate on this point in \S \ref{clr} 
\end{itemize}

It should be emphasized that our calculations are valid only when $k\ell$ is greater than unity. Consequently, in the soft limit of amplitudes, higher order $k$ corrections in flat spacetime are always dominant as compared to the $\f{1}{\ell}$ corrections due to the de Sitter background. This raises the question of whether a physical scenario exists when the perturbative effects due to the de Sitter potential are distinguishable or dominant compared to higher order $k$ corrections in flat spacetime. Indeed there is a class of asymptotic observables like memory effects where the $\f{1}{\ell}$ corrections are visible. For example, only the $\f{1}{k}$ mode contributes to the flat space electromagnetic velocity memory effect; the contribution of the subleading (in $k$) modes to this observable is at least exponentially suppressed. Now in the presence of de Sitter potential, the $\f{1}{k^3\ell^2}$-mode contributes as a power law term to the memory effect and hence is clearly distinguishable from flat space subleading (in $k$) soft modes \cite{2108}. Similar modifications are expected for asymptotic charges.

The rest of the paper is organized as follows. In \S \ref{scmode}, we find perturbative corrections to the scalar field modes in de Sitter spacetime. We demand that the corrections are such that the modes behave like plane waves in the limit $\ell \rightarrow \infty$ and are orthogonal up to $O\left(\f{1}{\ell^2}\right)$. We then construct the scalar propagator using these modes. We find similar corrections to gauge field modes in \S \ref{gf}. The LSZ prescription in perturbative de Sitter spacetime is derived in \S \ref{clsz}, and is used to define S-matrix in a small patch. In \S \ref{corrections}, we finally compute the leading and subleading corrections to the flat space soft photon theorem. Having calculated the corrections to the soft photon theorem, in \S \ref{clr}, we compare our results with results obtained from classical radiation analysis in \cite{2108}. We have checked the universality of our subleading ($O\left(\f{1}{\ell^2}\right)$) results by considering a different class of scalar modes in Appendix \ref{nmode}.

\section{Scalar field in de Sitter spacetime}\label{scmode}
In this section, we consider the solutions of the scalar field equation of motion in de Sitter spacetime. We will use stereographic coordinates \cite{stereo, stereo2}, which are well-defined throughout the static patch of de Sitter spacetime. In these coordinates, the metric takes the following form 
\begin{align}
g_{\mu\nu}&=\Omega^2 \eta_{\mu\nu} ,\ \ \Omega=\f{1}{1+x^2/4\ell^2}.\label{coord}
\end{align}
where $\ell$ is the curvature length of de Sitter spacetime and $x^2=\eta_{\mu\nu}x^\mu x^\nu$ where $\eta_{\mu\nu}$ is the $d$ dimensional Minkowski metric \footnote{We work with mostly positive signature i.e. $\eta_{\mu \nu} = \rm Diagonal \{-1,1, \ldots, 1\}$}. We will use Greek indices which run from $(0, \cdots , d-1)$ to denote de Sitter tensors.

 An advantage of using the above coordinate system is that the metric is conformally flat. Our aim is to keep the leading order correction in the limit when the curvature length ($\ell$) is large. In this limit, the metric takes the following form:
\begin{equation}
g_{\mu\nu}\approx \eta_{\mu\nu} -\f{x^2}{2\ell^2} \eta_{\mu\nu}.\label{metric}
\end{equation}
From the above expression, it is clear that the correction is perturbative as long as $x^\mu << \ell$ for every component '$\mu$'. Thus we will restrict to a region of size $R$ such that the points in $R$ have $x^\mu << \ell$ for every component '$\mu$'. Note that $x^2=-4\ell^2$ is a singular surface in this coordinate system, but this singularity does not affect our analysis as $|x^2|<<4\ell^2$ for us. 

\begin{figure}[h]
\centering
\includegraphics[width=0.62\textwidth]{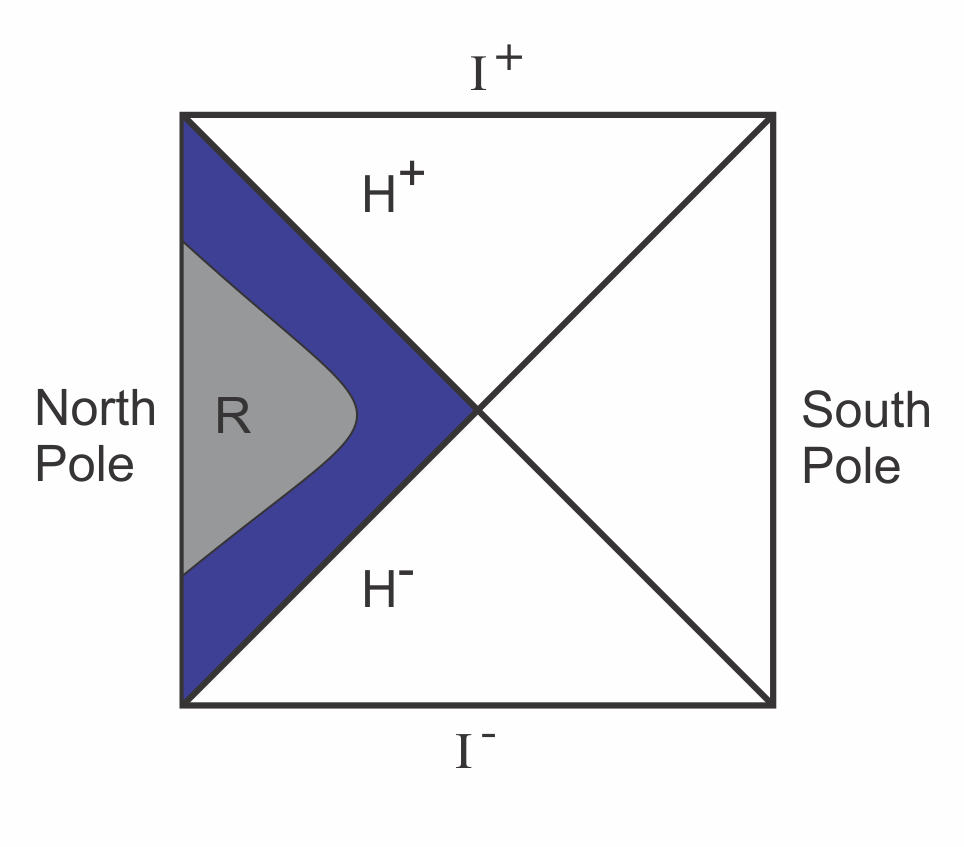}
\caption{Penrose diagram of the de Sitter spacetime  }
\label{penrose}
\end{figure}
The Penrose diagram of de Sitter spacetime is shown in Figure \ref{penrose}. The blue-coloured triangle in Figure \ref{penrose} is the static patch of de Sitter spacetime. In our setup, the scattering region, as well as the detector, lie within the region '$R$' in the static patch of the de Sitter spacetime.

\subsection{Modes of scalar fields}
In this subsection, we will write down the solutions of the free scalar field equation of motion in de Sitter spacetime whose metric is given in \eqref{metric}. The free scalar field equation is given by 
\be\label{scalareom}
[\nabla^2-m^2]\phi =0,\ee
where $\nabla^2$ is the D-Alembertian operator in de Sitter spacetime and $m$ is the mass of the scalar field. In the limit $\ell$ goes to infinity, the above equation takes the following form 
\be \label{sm}
[\nabla^2-m^2]\phi = \left(1+\f{x^2}{2\ell^2}\right)\Box\phi +\f{2-d}{2\ell^2}x.\p\phi-m^2\phi=0,\label{scalareom1}\ee
where we have kept $O(\f{1}{\ell^2})$ terms in the perturbation theory. 

In embedding space co-ordinates $(X^A)$, the solution of scalar field equation of motion given in \eqref{scalareom} is given by  \cite{Bros:1995js}:
\be 
g_{p}= {\mathcal{C}_\ell}\left(\f{X^A\xi^B\eta_{AB}}{\ell}+i\varepsilon\right)^{-\Delta},\label{XP}\ee
The index $A$ in the above expression takes values from 0 to $d$ where $d$ is the dimension of de Sitter spacetime and $\eta_{AB} = \rm Diagonal \{-1,1,1, \ldots, 1\}$ is the flat space metric in $d+1$ dimensional embedding space. $\mathcal{C}_\ell$ is a normalization constant which we will fix later.
The vector $\xi^A$ and the parameter $\Delta$ in \eqref{XP} have to satisfy following equations respectively : $\xi^A\xi^B\eta_{AB}=0$ and $$\Delta(\Delta - (d-1)) = -m^2 \ell^2.$$ Hence $\Delta$ can take two values
\be\label{Delta}
\Delta_{\mp} = \frac{1}{2} \left(d-1\mp \sqrt{(d-1)^2-4 \ell^2 m^2}\right)
\ee
As we will see below, the solution $\Delta_-$ corresponds to incoming modes ($e^{- i p x}$) and  $\Delta_+$ to outgoing modes ($e^{i p x}$) in the flat space limit.

 We parametrize $\xi^A=(\f{p^\mu}{m},1)$ such that $\eta_{\mu\nu}p^\mu p^\nu = - m^2$. Later $p^\mu$ will be identified with the flat space momentum. The embedding space coordinate $X^A$ can be expressed in terms of stereographic coordinates $x^\mu$ as
\be \label{emb}
 X^\mu=\f{x^\mu}{1+\f{x^2}{4\ell^2}}.
\ee
and the last component $X^d$ is fixed by the relation $\eta_{AB}X^A X^B = \ell^2$. 

Let us discuss the behaviour of the solution given in \eqref{XP} in the limit when $\ell$ is taken to infinity. The normalization factor $\mathcal{C}_\ell$ is chosen to get the right normalization in flat spacetime  limit i.e.
\be
\lim_{\ell \rightarrow \infty} g_p(x) \rightarrow \f{e^{ipx}}{\sqrt{2E_p}}\left (1+O\left(\f{1}{\ell}\right)\right )
\ee
We obtain: 
\be \label{C}\mathcal{C}_\ell =e^{-m\pi \ell + \frac{(d-1)i\pi}{2}+\f{(d-1)^2\pi}{8m\ell}}\Big[1 - \f{d (d - 1) (d - 2)}{24 \ell^2  m^2}\ \Big].
\ee
Substituting $\mathcal{C}_\ell$, $X_A$,  $\xi_A$ and $\Delta_+$ in \eqref{XP} and keeping upto $O(\f{1}{\ell^2})$ terms, we obtain
\begin{align}
g_p(x)&=\f{e^{ipx}}{\sqrt{2E_p}}\Big[1+(d-1)\f{p.x }{2\ell m}+\f{imx^2}{2\ell }+\f{i(p.x)^2}{2\ell m}- \f{d (d - 1) (d - 2)}{24 \ell^2  m^2}-(d-1)^2\f{ip.x}{8m^2\ell^2}\nn\\
&+(d-1)\f{x^2}{4\ell^2} +(d^2-1)\f{(p.x)^2}{8\ell^2m^2}+\f{d\ ip.x\ x^2}{4\ell^2}+(3d+1)\f{i(x.p)^3}{12m^2\ell^2}-\f{(p.x)^4}{8\ell^2m^2}-\f{(p.x)^2 x^2}{4\ell^2}-\f{m^2 x^4}{8\ell^2}\Big].\label{gp}
\end{align}
 It can be checked that $g_p$ solves \eqref{scalareom1} with the on-shell condition $$p^2=-E_p^2+{\vec{p}}^2=-m^2. $$
As discussed before, $g_p(x)$ is a valid solution only for $x^\mu < \ell$. 

Let us discuss the orthogonality of the modes $g_p(x)$. The usual Klein Gordon inner product is given by
$$(f_1,f_2)_{\rm KG}=-i\int d^{d-1}x\sqrt{-g}\ [\ f_1^*(t,\vec{x})\ \p^t f_2(t,\vec{x}) -f_2(t,\vec{x})\ \p^t f_1^*(t,\vec{x})\ ].$$
where we have chosen a constant time slice at time $t$ to define the inner product. 
In our case, the perturbative solutions given in \eqref{gp} are well defined only inside the region $R$ discussed earlier, i.e. for $x^\mu < R< \ell$. Since these solutions do not vanish on the boundary of $R$, we introduce an exponentially damping factor in the inner product to ensure that we do not get any boundary contributions. 
Therefore we define the inner product as follows
\be
\begin{split}
(f_1(x),f_2(x))=&-i\int d^{d-1}x\sqrt{-g}\ e^{-\epsilon \f{|\vec{x}|}{R}}\  [\ f_1^*(t,\vec{x})\ \p^t f_2(t,\vec{x}) -f_2(t,\vec{x})\ \p^t f_1^*(t,\vec{x})\ ].
\end{split}\label{in}
\ee
where $\epsilon$ is a small positive number which will be taken to zero at the end. We checked that $\p_t (f,g) =0$ as required. The modes given in \eqref{gp} are orthogonal with respect to the inner product defined in \eqref{in} i.e.
\be
(g_p,g_q)=  (2 \pi)^{d-1}\delta^3(\vec{p}-\vec{q}).
\ee

Using the orthogonal solutions $g_p$, we can now write down the mode expansion of the free scalar field as
\be
\phi(x)=\int \f{d^{d-1}p}{(2\pi)^{d-1}} \ [\ a_p\ g_p(x)\ +\ a^\dagger_p\ g^*_p (x)\ ].\nn
\ee
where $a_p^{\dagger}$ and $a_p$ are the usual creation and annihilation operators respectively which satisfy
\be\label{aad}
[ a^{\dagger}_p,  a_q] = \delta^{d-1}(p-q)
\ee
and can be expressed in terms of the modes $g_p$ as follows:
\be\label{ap}
a_p = (g_p(x), \phi(x)),  \qquad a_p^\dagger = (g^*_p(x), \phi(x))
\ee
where $(.,.)$ denotes the inner product defined in \eqref{in}.

\subsection{Feynman propagator}
In this section, we obtain $O(\f{1}{\ell^2})$ corrections to the flat space Feynman propagator. We do this by two methods: a) By explicitly solving the equation for the Greens function and b) by decomposing the Greens function in terms of the scalar field modes. Even though these two forms of the Greens function look different, we show that they are equivalent.

\subsubsection{Using equation of motion}
We start with the differential equation for the Feynman propagator :
\be\label{xeq}
[\nabla_x^2- m^2]\ D(x,y)= \left[\left(1+\f{x^2}{2\ell^2}\right)\Box_x-m^2+\f{(2-d)}{2\ell^2}x.\p\right]{D} =i \f{\delta^4(x-y)}{\sqrt{-g}}.\label{propeom}\ee
The propagator must also satisfy a similar equation in the variable $y$. Let us first solve the equation in variable $x$. The solution $D$ can be expanded as follows \cite{PhysRevD.20.2499, bunch2}
$${D}(x,y) = {D}_0(x,y)+ \delta D(x,y),$$
where
$${D}_0(x,y)=-i\int \f{d^4p}{(2\pi)^4}\f{e^{ip.(x-y)}}{p^2+m^2-i\epsilon}$$
is the flat space  Feynman propagator and $\delta {D}(x,y)$ denotes order $O \left(\f{1}{\ell}\right)$ corrections. Now $\delta D$ has to satisfy following equation 
\be \label{preq}
 [\Box_x-m^2]\delta D+\f{(2-d)}{2\ell^2}x.\p{D_0}+m^2\f{x^2}{2\ell^2}{D}_0=i\f{(d-2)}{4}\f{x^2}{\ell^2}\delta^4(x-y).\label{101}\ee
Substituting the Fourier representation of $\delta D$ i.e.
$$ \delta D (x,y)=-i\int \f{d^4p}{(2\pi)^4}e^{ip.(x-y)} \delta \widetilde{D}(p,y).$$
in \eqref{preq} and writing $x$ as $-i \partial_p (e^{i p x})$ and then using integration by parts in $p$ we obtain:
\be 
\begin{split}
-(p^2+m^2)\delta \widetilde{D}(p,y)&=-\f{4m^4}{\ell^2}\f{1}{(p^2+m^2)^3}-\f{2m^2(d-3-ip.y)}{\ell^2(p^2+m^2)^2}\\
&-\f{m^2y^2-(d-2)ip.y+(d-2)^2}{2\ell^2(p^2+m^2)} -\f{(d-2)}{4}\f{y^2}{\ell^2}.
\end{split}
\ee
Hence the full propagator is given by:
\begin{align} 
D(x,y)=-i\int\f{d^dp}{(2\pi)^d}\f{e^{ip.(x-y)}}{p^2+m^2}\Big[ & 1+\f{4m^4}{\ell^2(p^2+m^2)^3}+\f{2m^2(d-3-ip.y)}{\ell^2(p^2+m^2)^2}\nn\\
&+\f{m^2y^2-(d-2)ip.y+(d-2)^2}{2\ell^2(p^2+m^2)}+\f{(d-2)}{4}\f{y^2}{\ell^2}\Big]\label{fpasym}
\end{align}
By construction, $D(x,y)$ satisfies \eqref{propeom}. It can also be checked that $D(x,y)$ satisfies similar equation in the variable $y$. Though the above form of the propagator is not explicitly symmetric in $x$ and $y$ it can be brought to an explicitly symmetric form using integration by parts. We will derive the symmetric form of the propagator in the following subsection.

\subsubsection{Symmetric form of Feynman propagator}\label{fnp}
In this subsection, we will use the solutions $g_p$ of \eqref{scalareom1} to derive a symmetric form of the Feynman propagator. Given an orthogonal basis of solutions $g_p$, the Greens function (which is the Feynman propagator with the correct $i \epsilon$ prescription) can be written by: 
\be
D(x,y) = -i\int\f{d^d p}{(2\pi)^d}\f{g_p(x)g_p^*(y)}{p^2+m^2-i\epsilon}
\ee
It is easy to check that $D(x,y)$ satisfies \eqref{xeq} and a similar equation in $y$ variable. Using $g_p$ from \eqref{gp} and keeping terms upto $O\left(\f{1}{\ell^2}\right)$, we obtain
\be
 D(x,y) =\  D^{(0)}+ D^{(1)} + D^{(2)}.
 \ee
where $D^{(0)}$ is the flat space  propagator, $ D^{(1)}$ and $D^{(2)}$ are $O\left(\f{1}{\ell}\right)$ and $O\left(\f{1}{\ell^2}\right)$ contributions respectively. They are given by:
\begin{align}\label{gs12}
{D}^{(0)}&=\int \f{d^d p}{(2\pi)^d}\f{e^{ip.(x-y)}}{p^2+m^2},\nn\\
 D^{(1)}&=    \int \f{d^d p}{(2\pi)^d}\f{e^{ip.(x-y)}}{p^2+m^2}\ \Big[ \frac{(d-1) p.x}{2 \ell m}+\frac{(d-1) p.y}{2 \ell m}+\frac{i (p.x)^2}{2 \ell m}-\frac{i (p.y)^2}{2 \ell m}+\frac{i m x^2}{2 \ell}-\frac{i m y^2}{2 \ell}\Big]=\ 0,\nn\\
 D^{(2)}&=    \int \f{d^d p}{(2\pi)^d}\f{e^{ip.(x-y)}}{p^2+m^2}\ \Big[ \frac{\left(d^2-1\right) (p.x)^2}{8 \ell^2 m^2}+\frac{\left(d^2-1\right) (p.y)^2}{8 \ell^2 m^2}+\frac{i (d-1) (p.x)^2 p.y}{4 \ell^2 m^2}+\frac{(d-1)^2 p.x p.y}{4 \ell^2 m^2}\nn\\
 &-\frac{i (d-1) p.x (p.y)^2}{4 \ell^2 m^2}+\frac{i (3 d+1) (p.x)^3}{12 \ell^2 m^2}-\frac{i (d-1)^2 p.x}{8 \ell^2 m^2}+\frac{i (d-1)^2 p.y}{8 \ell^2 m^2}-\frac{i (3 d+1) (p.y)^3}{12 \ell^2 m^2}\nn\\
 &+\frac{i (d-1) x^2 p.y}{4 \ell^2}+\frac{i d x^2 p.x}{4 \ell^2}-\frac{i (d-1) y^2 p.x}{4 \ell^2}-\frac{i d y^2 p.y}{4 \ell^2}+\frac{(d-1) x^2}{4 \ell^2}+\frac{(d-1) y^2}{4 \ell^2}+\frac{(p.x)^2 (p.y)^2}{4 \ell^2 m^2}\nn\\
 &-\frac{(p.x)^4}{8 \ell^2 m^2}-\frac{(p.y)^4}{8 \ell^2 m^2}-\frac{m^2 x^4}{8 \ell^2}+\frac{m^2 x^2 y^2}{4 \ell^2}-\frac{m^2 y^4}{8 \ell^2}+\frac{x^2 (p.y)^2}{4 \ell^2}-\frac{x^2 (p.x)^2}{4 \ell^2}+\frac{y^2 (p.x)^2}{4 \ell^2}-\frac{y^2 (p.y)^2}{4 \ell^2}\Big]
\end{align}
The above expression of $D(x,y)$ is manifestly symmetric in $x$ and $y$. Let us now compare the above propagator with the asymmetric propagator \eqref{fpasym}, which was derived in the previous subsection by explicitly solving the equation of motion. 
\begin{itemize}
    \item The zeroth order propagator i.e. the $\ell$ independent part of \eqref{fpasym} is same as $D^{(0)}$.
    \item 
    We can express the variable $x$ appearing in $D^{(1)}$ as $-i \partial_p (e^{i p x})$ and then do integration by parts in $p$, hence converting it into a $y$ dependent piece. One can check that after doing these manipulations, $D^{(1)}$ vanishes. This is consistent with the absence of the $\f{1}{\ell}$-piece in \eqref{fpasym}.
    \item The $\f{1}{\ell^2}$ contribution of both the answers naively look very different, but again, by using integration by parts in $p$, we can obtain an expression for $D^{(2)}$ in terms of variable $y$ only. We checked that this matches the $\f{1}{\ell^2}$-piece of $D(x,y)$ derived in the previous subsection.
\end{itemize}
\section{Gauge field in de Sitter spacetime}\label{gf}
In this section, we find the mode expansion and propagator of the U(1) gauge field in de Sitter spacetime. The equation of motion of the U(1) field in stereographic coordinates to $O\left(\f{1}{\ell^2}\right)$ is given by
\be
\left(1+\frac{x^2}{2\ell^2}\right)\left(\Box A_\mu - \p_\mu(\p^\nu A_\nu)\ \right)\ +\ \f{(4-d)}{2\ell^2}[\ x.\p A_\mu -\p_\mu(x^\nu A_\nu)+A_\mu]\ =\ j_\mu.
\ee
where $j_\mu$ is the U(1) matter current and we have only kept $O(\f{1}{\ell^2})$ corrections to flat space equation of motion. Let us fix the gauge by setting
\be \label{gauge1}\p^\nu A_\nu
+\f{(4-d)}{2\ell^2}\ (x^\nu A_\nu)=0 \ee
Notice that this is not a covariant gauge choice, but it simplifies the equation of motion. Using the gauge condition \eqref{gauge1} in the equation of motion, we obtain
\be\label{geom}
\Box A_\mu \ +\ \f{(4-d)}{2\ell^2}\ [x.\p A_\mu +A_\mu]\ =\ j_\mu \left(1-\frac{x^2}{2\ell^2}\right).
\ee
For $d=4$ and $j_\mu = 0$, the above equation reduces to the flat space equation of motion, and the gauge choice becomes the standard Lorentz gauge. This happens because the de Sitter metric \eqref{metric} is related to the flat space metric by Weyl transformation and the pure Maxwell Lagrangian is Weyl invariant in $d=4$; hence we can perform Weyl transformation on the de Sitter metric to obtain the flat space equation of motion.

The equation of motion \eqref{geom} admits homogenous solutions (i.e. $j_\mu = 0$) of the form 
\be \label{hk} f^h_{k\mu}(x) =\ \frac{1}{\sqrt{2 E_k}} \varepsilon_\mu^{h}(k)\ \left(1+\f{(d-4)}{8}\frac{x^2}{\ell^2}\right)\ e^{ik.x},\ \ k^2=\f{(d-2)(d-4)}{4\ell^2}, \label{Amodes} \ee
where $E_k$ is the zeroth component of $k^\mu$. In $d\neq 4$, $k^2$ is non-zero, which is expected for generic curved spacetime. As discussed earlier, $d=4$ is a special case and the solution reduces to plane waves with $k^2 = 0$ in four spacetime dimensions. After fixing all redundant degrees of freedom of the gauge field, one is left with $d-2$ physical degrees of freedom. Here we have used $h$ to denote these physical helicity states with $h$ taking values from 1 to $d-2$. 

The Maxwell inner product on the solution space is defined as
\begin{align}
(g^{\mu\nu}\ f^h_{k\mu},f^{h'}_{k'\nu})=&-i\int d^{d-1}x\sqrt{-g}\ g^{\mu\nu}(x)\ e^{-\epsilon \f{|\vec{x}|}{R}}\ [\ f^{*h}_{k\mu}(t,\vec{x})\ \p^t f^{h'}_{k'\nu}(t,\vec{x}) -f^{h'}_{k'\mu}(t,\vec{x})\ \p^t f^{*h}_{k\nu}(t,\vec{x})\ ]\nn\\
=&\ {(2\pi)^{d-1}} \delta_{h,h'}\ \delta^3(\vec{k}-\vec{k'}).\label{inA}
\end{align}
where we have again added exponential damping factor to get rid of the boundary terms. Hence the gauge field can be expanded in terms of the above modes as follows
\be
A_\mu(x)=\sum_{h=1}^{d-2}\int \f{d^{d-1}k}{(2\pi)^{d-1}} \ [\  a^h_k\ f_{k\mu}^{h}(x)\ +\  a^{h\dagger}_k \  f_{k\mu}^{*h}(x)\ ].\nn
\ee
Let us write down the implication of the gauge condition \eqref{gauge1} in momentum space. We get 
\be ik^\nu\varepsilon^h_\nu\  a^h_k\ e^{ik.x}
+\f{(4-d)}{4\ell^2}\  \varepsilon^h_\mu x^\mu\ a^h_k \ e^{ik.x}=0 \label{gauge}
\ee
Thus $\varepsilon.k\neq 0$ in general.

Next, we derive the Feynman propagator of the gauge field. The propagator ($ G_{\mu\nu'}$) has to satisfy the following equation of motion in the first argument $x$ \cite{Poisson:2011nh}
\be\label{gphoton}\left[1+\f{x^2}{2\ell^2}\right]\Big[\Box_x +\ \f{(4-d)}{2\ell^2}\ [x.\p +1]\ \Big] G_{\mu\nu'}(x,x') =ig_{\mu\nu'}(x,x')\f{\delta^d(x-x')}{\sqrt{-g}}.\ee
The gauge field propagator $G_{\mu\nu'}(x,x')$ is a bivector defined at points and $x$ and $x'$ and $g_{\mu\nu'}(x,x')$ is the parallel propagator along the geodesic connecting $x$ and $x'$. Hence $g_{\mu\nu'}(x,x)$ is equal to the metric $g_{\mu\nu}(x)$. So we obtain
\be
\Big[\Box +\ \f{(4-d)}{2\ell^2}\ [x.\p +1]\ \Big] G_{\mu\nu'}(x,x') =i\eta_{\mu\nu'}\delta^d(x-x')\left[1+\f{(d-4)}{4}\f{x^2}{\ell^2}\right].\label{photonG}
\ee
Thus in $d= 4$, the propagator equation of motion becomes identical to the flat space case i.e. 
$$\Box G_{\mu\nu'}(x,x')= i\eta_{\mu\nu'}\delta^d(x-x').$$
The solution to the propagator is given by 
$$ G_{\mu\nu'}(x,y) =-i\eta_{\mu\nu'}\ \int \f{d^d k}{(2\pi)^d}\f{e^{ik.(x-y)}}{k^2} \Big[1+\f{(d-4)}{2\ell^2}\frac{(d-3)}{k^2}-\f{(d-4)}{2}\frac{ik.y}{\ell^2}+\f{(d-4)}{4}\frac{y^2}{\ell^2}\Big].$$
The propagator also admits a symmetric form in variables $x$ and $y$ with the following decomposition in terms of the modes given in \eqref{Amodes}:
$$ G_{\mu\nu'}(x,y) =-i\eta_{\mu\nu}\ \int \f{d^d k}{(2\pi)^d}\f{e^{ik.(x-y)}}{k^2-\f{(d-2)(d-4)}{4\ell^2}} \Big[1+\f{(d-4)}{8}\frac{x^2}{\ell^2}+\f{(d-4)}{8}\frac{y^2}{\ell^2}\Big].$$
Using integration by parts, similar to the previous section, we checked that the above two forms of propagator are equivalent.

\section{Perturbative S-matrix in de Sitter spacetime}\label{clsz}
In this section, we will define a perturbative S-matrix for de Sitter spacetime. In flat spacetime, the S-matrix for a process involving the scattering of a set of incoming plane waves with momenta $p_i$ to a set of outgoing plane waves with momenta $p_j$, is defined as 
\be
S_{\rm flat}(\lbrace p_i, p_j \rbrace)=\lim_{t\rightarrow\infty}\langle \ \mathcal{T}\ \prod_{j\in \text{out}}\sqrt{2E_{p_j}}\textbf{a}_{p_j}(t)\ \prod_{i\in \text{in}}\sqrt{2E_{p_i}}\textbf{a}^\dagger_{p_i}(-t) \ \rangle\ .\label{Sf}
\ee  
Here $\textbf{a}_{p_j} (\textbf{a}_{p_i}^\dagger)$ are the annihilation (creation) operators for plane waves, and $\mathcal{T}$ represents time-ordering.

It is well-known that a globally defined S-matrix might not exist in an arbitrary curved spacetime. Nonetheless, it should be possible to define the S-matrix in a local patch where the effects of the background can be treated perturbatively. We will use such a local construction of S-matrix \cite{S} in the region $R$ inside the static patch defined in \S \ref{scmode}.  We can associate the following S-matrix with the scattering of a set of scalar modes $g_{p_i}$ (or gauge field modes $f^h_{k\mu}$) into another set of modes carrying momentum $p_j$
\be
\Gamma(\lbrace p_i, p_j \rbrace) =\lim_{t\rightarrow T}\ \langle \ \mathcal{T}\ \prod_{j\in \text{out}}\sqrt{2E_{p_j}}a_{p_j}(t)\ \prod_{i\in \text{in}}\sqrt{2E_{p_i}}a^\dagger_{p_i}(-t)\ \rangle\ .\label{S}
\ee  
Here the creation and annihilation operators are inserted at the boundaries of the region $R$ inside the static patch. It is assumed that $T$ is sufficiently larger than the interaction time scale.

We can express the creation and annihilation operators appearing above in terms of field operators by using the inner product defined in \eqref{in}. For the annihilation operator corresponding to the scalar field, we have
\begin{align}
a_p(T)-a_p(-T)&=\int^T_{-T} dt\ \p_t a_p\nn\\
&=-i\int^T_{-T} dt d^{d-1}x\ e^{-\epsilon \f{|\vec{x}|}{R}}\ \p_t  \left[\ \sqrt{-g}\ [\ g_p^*\p^t \phi -\phi\p^t g_p^*\ ]\ \right] \nn\\
&=-i\int^T_{-T} dt d^{d-1}x\ e^{-\epsilon \f{|\vec{x}|}{R}}\ \p_\mu\big[\sqrt{-g}\ [\ g_p^*\p^\mu \phi -\phi\p^\mu g_p^*\ ]\ \big]\nn\\
&=-i\int^T_{-T} dt d^{d-1}x\ e^{-\epsilon \f{|\vec{x}|}{R}}\ \sqrt{-g}\ \nabla_\mu [\ g_p^*\nabla^\mu \phi -\phi\nabla^\mu g_p^*\ ]\ 
\end{align}
where we have used \eqref{in} and \eqref{ap} in the first step. In the third step, we added a boundary term. Using the fact $g_p^*$ satisfies the scalar equation of motion given in \eqref{scalareom} and taking $\epsilon \rightarrow 0$, we get
\begin{align}
a_p(T)-a_p(-T)&=-i\int^T_{-T} dt d^{d-1}x\ \sqrt{-g}\  g_p^*\ [\nabla^2 - m^2]\ \phi .
\end{align}
We derive a similar formula for annihilation operators of the gauge field in Appendix \ref{gaugeap}. We obtain using \eqref{lszA} 
\begin{align}
a^h_k(T)-a^h_k(-T)&=-i\ \int^T_{-T} d^{d}x\ \sqrt{-g(x)}\ f^{*h}_{k \sigma}(x)\ \eta^{\sigma\mu}\ \mathcal{D}_x\ A_\mu(x) ,\end{align}
where $\mathcal{D}_x$ is defined in \eqref{Dx}.
Using the above expression in \eqref{S}, we get a curved space generalization of the LSZ formula for the S-matrix, which is given by
\begin{align}
\Gamma(\lbrace p_i,p_k\}, \{p_j \rbrace)&=\int \prod_{i,k\in \text{in}}\prod_{j\in \text{out}}\ [d^dx_i] [d^dy_j] [d^dz_k] \  \eta^{\sigma\mu} g_{p_j}(y_j)g^*_{p_i}(x_i)f^{*h}_{p_k \sigma}(z_k)\ \nn\\
& (-i)[\nabla_i^2 - m^2]\  (-i)[\nabla_j^2 - m^2]\  (-i)\mathcal{D}_{z_k} \langle \mathcal{T}\ A_{\mu}(z_k)\phi({x_i})\ldots \phi({y_j})\rangle .\label{LSZ}
\end{align} 
where the measure factors are given by:
\be
[d^dx_i] [d^dy_j] [d^dz_k] 
 = d^dx_i\ d^dy_j\ d^dz_k \ \sqrt{2E_{p_i} (2E_{p_j})\ (2E_{p_k})} \ \sqrt{-g(x_i) \ (-g(y_j)) \ (-g(z_k))} \nn 
\ee
In the above formula, photons are taken to be incoming. We can write a similar formula for outgoing photons by just replacing $f^{*h}_{k\sigma}$ with $f^{h }_{k\sigma}$.

\section{Corrections to flat space soft photon theorems}\label{corrections}
In the previous sections, we found $O(\f{1}{\ell^2})$ corrections to the modes and propagators of both scalar fields and gauge fields. We are now set to compute the corrections to the flat space soft photon theorems in $d \geq 4$. The action for a complex scalar field minimally coupled to the U(1) gauge field is given by
\begin{equation}
\mathcal{S} = -\int d^{d}x \sqrt{-g}\ \Big[\frac{1}{4}g^{\mu\rho}g^{\nu\sigma}F_{\mu\nu}F_{\rho\sigma} + g^{\mu\nu}\left(D_{\mu}\phi\right)^{*}\left(D_{\nu}\phi\right) +\ V[|\phi|^2]\  \Big],
\end{equation}
where
$$D_\mu\phi=\p_\mu\phi-ieA_\mu\phi$$
and $V[|\phi|^2]$ is a gauge invariant scalar potential. The leading order (in $(\f{1}{k})$) soft photon theorem in flat spacetime relates an amplitude with $n$ hard particles and one soft photon to the amplitude without the soft photon. It takes the following form
\be
\Gamma_{n + 1}(\{p_i\}, k) = S(p_i , k) \Gamma_{n}(\{p_i\}),
\ee
where $p_i$ are the momentas of hard particles, $k$ is the soft momentum and $S(p_i , k)$ is the soft factor.

Due to the presence of de Sitter background, the flat space soft factor as well as the amplitude $\Gamma_{n}(\{p_i\})$ receives $O(\f{1}{\ell})$ corrections i.e. 
\begin{align}
    \label{notation}
\Gamma_{n + 1}(\{p_1...p_n\}, k) = \left(S^{(0)} + \f{1}{\ell} S^{(1)} + \f{1}{\ell^2} S^{(2)} \right) \left(\Gamma^{(0)}_{n}(\{p_i\}) + \f{1}{\ell} \Gamma^{(1)}_{n}(\{p_i\})+ \f{1}{\ell^2} \Gamma^{(2)}_{n}(\{p_i\})\right)
\end{align}
where $S^{(0)}$ is the flat space soft factor.

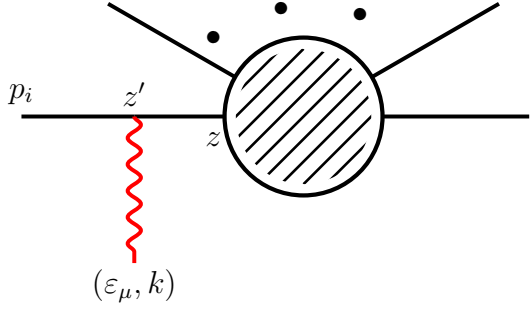
\begin{figure}[h]
\begin{center}
\begin{tikzpicture}[line width=1.5 pt, scale=1.5]

\begin{scope}[shift={(6,0)}]

\draw[realscalar] (.7,0)--(2,0);

\draw[realscalar][rotate=30] (.7,0)--(2,0);

\draw[realscalar][rotate=180] (.7,0)--(1.5,0);

\draw[realscalar][rotate=180] (2.5,0)--(1.5,0);
\node at (-0.8,.7) {$\bullet$};
\node at (-0.2,.95) {$\bullet$};
\node at (0.5,.9) {$\bullet$};
\node at (-1.5,.2) {$z'$};
\node at (-.8,-.2) {$z$};
\draw[realscalar][rotate=150] (.7,0)--(2,0);

\draw[photon][rotate=180] (1.5,0)--(1.5,1.3);

\node at (-1.5,-1.5) {$(\varepsilon_{\mu }, k)$};

\node at (-2.5,.2) {$p_i$};

\begin{scope}[shift={(0,0)}, scale=2]
	\draw [ultra thick] (0,0) circle (.35);
	\clip (0,0) circle (.3cm);
	\foreach \x in {-.9,-.8,...,.3}
	\draw[line width=1 pt] (\x,-.3) -- (\x+.6,.3);
\end{scope}
 
\end{scope}

\end{tikzpicture}
\end{center}
\caption{Soft photon theorem}
\label{fig1}
\end{figure}

Let us now compute $\Gamma_{n + 1}(\{p_i\}, k)$ using the curved space LSZ formula discussed in \eqref{LSZ} of the previous section. We consider the diagram in Fig \ref{fig1} where the soft photon is attached to the external particle carrying momentum $p_i$ via the minimal three-point coupling at point $z'$. The diagrams in which the external soft photon is attached to an internal leg or to four (or higher) point coupling start contributing at $O(k^0)$; hence are not relevant for our analysis. In the rest of this paper, we will work with the convention that all the external momenta are outgoing. We will suppress the helicity index $h$ of the gauge field modes for notational brevity.

\begin{align}
\Gamma_{n + 1}(\{p_1...p_n\}, k)\ &= \sum_{i = 1}^{n} e_i\int d^d z_i \sqrt{-g(z_i)}\ g^*_{p_i}(z_i)(-i) (\nabla_{z_i}^2 - m^2) ~ \int d^d y \sqrt{-g(y)}\ f^{* }_{k\rho}(y) \eta^{\rho \sigma} \ (-i)(\mathcal{D}_y)\nn\\
&~ \int \prod_{\substack{j = 1\\ j \neq i}}^{n-1} d^d z_j \sqrt{-g(z_j)}\ g^*_{p_j}(z_j)(-i) (\nabla_{z_j}^2 - m^2)\int d^d z \sqrt{-g(z)} \int d^d z' \sqrt{-g(z')}\nn\\
& \langle \mathcal{T}~ \phi(z_i)\  A_\sigma(y)\ g^{\mu\nu}(z')\ A_\mu(z')\ [\phi^*(z')\p_\nu\phi(z')-\phi(z')\p_\nu\phi^*(z')]\ G[\phi(z)...\phi(z_j)]\rangle\nn
\end{align} 
where $G[\phi(z)...\phi(z_j)]$ is the n-point scalar correlator which receives contributions from any arbitrary interaction vertices of the theory. Performing Wick contractions between various fields, we obtain the corresponding Greens functions and subsequently use the equations of motion of the Greens functions (from \eqref{xeq} and \eqref{G11}) to get
\begin{align}
\Gamma_{n + 1}(\{p_1...p_n\}, k)\ &= (-i)^{n}\sum_{i = 1}^{n} e_i\int d^d z_i \ g^*_{p_i}(z_i) \int \prod_{\substack{j = 1\\ j \neq i}}^{n-1} d^d z_j \sqrt{-g(z_j)}\ g^*_{p_j}(z_j) (\nabla_{z_j}^2 - m^2)\nn\\
&\times \int d^d z \sqrt{-g(z)} ~\int d^d z' \sqrt{-g(z')}\ 
  f^{*}_{k\mu}(z')\  g^{\mu \nu}(z')\  \nn\\
 &\times \left[i\delta(z_i-z')\p'_\nu D(z',z)- \sqrt{-g(z_i)}D(z',z)\p'_\nu \left(\f{i\delta(z_i-z')}{\sqrt{-g(z')}}\right)\right] \ G[\{\phi(z_j)\}].\nn
 \end{align}  
Keeping terms only up to $O\left(\f{1}{\ell^2}\right)$, the second term can be simplified as
\begin{align}
  &\sqrt{-g(z')}\ 
 f^{*}_{k\mu}(z') g^{\mu \nu}(z')\sqrt{-g(z_i)}D(z',z)\p'_\nu \left(\f{i\delta(z_i-z')}{\sqrt{-g(z')}}\right) \nn\\
 &= \frac{d}{4 \ell^2}f^{*}_{k\mu}(z')  z'^\mu D(z',z)i\delta(z_i-z') - \sqrt{-g(z')}\ 
 f^{*}_{k\mu}(z') g^{\mu \nu}(z')i\delta(z_i-z')\ [\p'_\nu - ik_\nu]\  D(z',z) \nn   
\end{align}
where we have done integration by parts to obtain the last piece. The $k_\nu$ term appears when the derivative acts on exponential in $f^{*\sigma}_k(z')$. This term is a pure gauge term in flat spacetime. In our case, we use the corrected gauge condition given in \eqref{gauge} to evaluate this term. We can now perform the $z_i$ integral using the Dirac delta function. We finally obtain
\begin{align}\label{softf}
\Gamma_{n + 1}(\{p_1...p_n\}, k)\ &=  -(-i)^{n+1} \int d^dz\ d^dz' \left(1 - \frac{d z^2}{4\ell^2}- \frac{(d-2) z'^2}{4\ell^2}\right)  \sum_{i = 1}^{n}e_i\ g^*_{p_i}(z') f^{*}_{k\mu}(z')\ \eta^{\mu\nu}\nn\\
 &   \left( 2\p'_\nu D(z',z) \   -\frac{z'_{\nu}}{\ell^2}D(z',z)\right) \ \int \prod_{\substack{j = 1\\ j \neq i}}^{n-1}d^d z_j \sqrt{-g(z_j)}\ g^*_{p_j}(z_j))
 (\nabla_{z_j}^2 - m^2)G[\{\phi(z_j)\}]\ \nn\\
 \end{align}

\subsection{Leading corrections to the flat spacetime  soft photon theorem}\label{ldg}
 In this section, we will compute the leading (i.e. $O\left(\f{1}{\ell}\right)$) corrections to the Feynman diagram shown in Fig \ref{fig1}. Notice that only scalar modes $g_p$ receive corrections at  $O\left(\f{1}{\ell}\right)$\footnote{We have already shown in subsection \ref{fnp} that $\frac{1}{\ell}$ corrections to scalar propagator vanishes after integration by parts.}. Using \eqref{gp}, \eqref{hk} and \eqref{gs12} in \eqref{softf}, we obtain
 \begin{align}\label{1ls}
\Gamma_{n + 1}(\{p_1...p_n\}, k)\ &=- (-i)^2\sum_{i = 1}^{n}e_i \int d^dz\ d^dz'  \Big[1+(d-1)\f{p_i.z' }{2\ell m}-\f{i m z'^2}{2\ell }-\f{i(p_i.z')^2}{2\ell m}\Big] e^{- i p_i z'}e^{-i k z'} \nn\\
& \times  \int \f{d^d p}{(2 \pi)^d}\left( 2  \f{\varepsilon.p}{p^2 + m^2} e^{i p (z'-z)} \right) \ \int \prod_{\substack{j = 1\\ j \neq i}}^{n-1}d^d z_j \ g^*_{p_j}(z_j) (-i)(\nabla_{z_j}^2 - m^2) G[\{\phi(z_j)\}]
\end{align}  
 The $\ell$ independent piece reproduces flat space soft factor as expected. This can be seen by noting that $z'$ integral in the corresponding piece gives $\delta^d (p -p_i-k)$, which can be used to perform the $p$ integral to obtain
\begin{align}
\Gamma^{(0)}_{n + 1}(\{p_1...p_n\}, k)&= \sum_{i = 1}^{n} e_i \frac{\varepsilon . p_i}{p_i.k}\int d^dz \ e^{-i(p_i + k) z}\ \int \prod_{\substack{j = 1\\ j \neq i}}^{n-1}d^d z_j \ g^*_{p_j}(z_j) (-i)(\nabla_{z_j}^2 - m^2) G[\{\phi(z_j)\}]\nn
\end{align}
We have used $k^2 = 0$ and $\varepsilon.k =0$, which are true at the $\ell^0$ order. Let us take the soft limit of this amplitude, i.e. $k\rightarrow 0$. For the leading order term, we will just set $k=0$ in the exponential. Since $p_i$ in the exponential is onshell, $e^{-ip_i  z}$ is just the factor obtained after performing LSZ on the $i^{th}$ leg. Thus we get
\begin{align}
\Gamma^{(0)}_{n + 1}(\{p_1...p_n\}, k)&= \sum_{i = 1}^{n}  e_i \frac{\varepsilon . p_i}{p_i.k} \Gamma^{(0)}_n(\{p_1...p_n\}) \ + \ O(k^0)\label{S0}
\end{align}
where $\Gamma^{(0)}_{n + 1}$ and  $\Gamma^{(0)}_{n }$ denotes the flat space contribution to S-matrix with $n+1$ and $n$ particles respectively.  Using the notation from \eqref{notation}, we get:
\be\label{flats}
S^{(0)} =   \sum_{i = 1}^{n} e_i\frac{\varepsilon . p_i}{p_i.k} + O(k^0)
\ee

Let us now analyse $\f{1}{\ell}$ pieces of \eqref{1ls}. To simplify $z'$ dependent pieces, we replace $z'$ by $-i \p_{p}(e^{i p z'})$, do integration by parts in $p$ and then perform the $z'$ integral. As an example, let us explicitly show the manipulations for one of the terms. \\
Let us take the following piece:
\begin{align}
&\int d^dz\ d^dz'  \Big[(d-1)\f{p_i.z' }{2\ell m}\Big] e^{- i p_i z'}e^{-i k z'} \int \f{d^d p}{(2 \pi)^d}\left( 2  \f{\varepsilon.p}{p^2 + m^2} e^{i p (z'-z)} \right) \nn\\
&= - i\f{d-1}{2 \ell m}\int d^dz\ d^dz' \ e^{- i (p_i + k)z'} \int \f{d^d p}{(2 \pi)^d}   \left( 2 \f{\varepsilon.p}{p^2 + m^2} e^{-i p z} \right)p_i.\p_{p}(e^{i p z'}) \nn\\
&=  i\f{d-1}{2 \ell m}\int d^dz  \int \f{d^d p}{(2 \pi)^d}   p_i.\p_{p}\left( 2  \f{\varepsilon.p}{p^2 + m^2} e^{-i p z} \right)\delta^d(p-p_i -k)\nn\\
&=  2 i\f{d-1}{2 \ell m} \int d^dz  \left(\frac{m^2 \varepsilon .p_i}{2 (k.p_i)^2}- i\frac{ \varepsilon .p_i}{2 k.p_i}z.p_i\right)e^{-i (p_i + k) z}
\end{align}  
where we used $p_i^2 = - m^2$ in the last step. Similarly analyzing all the terms in \eqref{1ls}, we finally obtain
\begin{align}\label{intpart}
\Gamma^{(1)}_{n + 1}(\{p_1...p_n\}, k) &= \sum_{i = 1}^{n}2  e_i\int d^dz \ e^{-i (p_i + k) z}\ 
\left[\f{\varepsilon.p_i}{2 p_i.k}\left((d-1)\frac{ p_i.z }{2 \ell m }-\frac{i m z^2 }{2 \ell }-\frac{i (z.p_i)^2 }{2 \ell m }\right)\right.\nn\\
&\left. +\frac{ m z.\varepsilon }{2 \ell k.p_i} -\frac{ m  \varepsilon .p_i \ k.z}{2 \ell (k.p_i)^2}\right]\int \prod_{\substack{j = 1\\ j \neq i}}^{n-1}d^d z_j \ g^*_{p_j}(z_j) (-i)(\nabla_{z_j}^2 - m^2) G[\{\phi(z_j)\}] + O(k^0)\nn\\
&=-i \f{ m}{\ell} \sum_{i = 1}^{n} e_i\left(\varepsilon .p_i\frac{k. \p_{p_i}}{ (k.p_i)^2}-  \frac{\varepsilon. \p_{p_i} }{ k.p_i} \right)\Gamma_n^{(0)}(\{p_i\})  + \frac{1}{\ell}S^{(0)} \Gamma_n^{(1)}(\{p_i\}) 
\end{align}
where $S^{(0)}$ is given in \eqref{flats} and $\Gamma_n^{(1)}(\{p_i\}$ contains $\f{1}{\ell}$ corrections to the $n$-particle scattering amplitude. In the last step, we have replaced $z$ with $i\p_{p_i}( e^{-i p_i z})$. Comparing with \eqref{notation}, we find that the leading correction to flat space soft factor is given by
\be\label{S1}
S^{(1)} =  -im \sum_{i = 1}^{n} e_i\left(\varepsilon .p_i\frac{k. \p_{p_i}}{ (k.p_i)^2}-  \frac{\varepsilon. \p_{p_i} }{ k.p_i} \right) + O(k^0).
\ee
\subsubsection{Consistency checks}\label{consist}
\begin{itemize}
\item Let us check how $S^{(1)}$ behaves under gauge transformations. At $O\left(\f{1}{\ell}\right)$, the gauge field modes do not receive any corrections. Hence the gauge transformation can be performed by shifting $\varepsilon_\mu \rightarrow \varepsilon_\mu - i k_\mu$. By performing this transformation, we conclude that $S^{(1)}$ is gauge invariant.
\item In Appendix \ref{nmode}, we construct a different class of scalar modes given in \eqref{fp} which behave like plane waves in $\ell \rightarrow \infty$ limit. We study the S-matrix elements corresponding to the scattering of these modes and compute the corrections to the flat space soft photon theorem. These modes do not have any $O\left(\f{1}{\ell}\right)$ piece. Hence there are no leading corrections to the flat space soft factor. Thus the $O\left(\f{1}{\ell}\right)$-mode is sensitive to the nature of the modes that are being scattered and hence it is non-universal.
\end{itemize}

\subsection{Subleading corrections to the flat space soft photon theorem} \label{5.22}
In this subsection, we compute subleading, i.e. $O\left(\f{1}{\ell^2}\right)$ corrections to flat space  soft photon theorem. At this order, the soft theorem gets corrections from the scalar modes, gauge field modes, scalar propagator, and the metric determinants. Re-writing \eqref{softf} as follows:
\begin{align}\label{l21}
\Gamma_{n + 1}(\{p_1...p_n\}, k)\ &=  (-i)^{n-1}\int d^dz\ d^dz' \left(1 - \frac{d z^2}{4\ell^2}- \frac{(d-2) z'^2}{4\ell^2}\right) \sum_{i = 1}^{n}e_i g^{*}_{p_i}(z') f^{*h \ \nu}_{k}(z') \nn\\
& \times \left( 2\p'_\nu D(z',z) \   -\frac{z'_{\nu}}{ \ell^2}D(z',z)\right)\mathcal{V}(\{p_j\})
\end{align} 
where 
$$\mathcal{V}(\{p_j\}) = \int \prod_{\substack{j = 1\\ j \neq i}}^{n-1}d^d z_j \sqrt{-g(z_j)}\ g^*_{p_j}(z_j) (-i)(\nabla_{z_j}^2 - m^2) \ G[\{\phi(z_j)\}]$$
There are three types of terms in \eqref{l21}, ones in which the factors multiplying exponential piece are only $z$ dependent, the second ones in which they are just $z'$ dependent and the last ones which are independent of both $z$ and $z'$. Let us study the last kind of terms below i.e.
\begin{align}\label{l22}
   (-i)^{n-1}\sum_{i = 1}^{n}e_i\int \prod_{\substack{j = 1\\ j \neq i}}^{n-1}d^d z_j e^{-ip_j.z_j} (\nabla_{z_j}^2 - m^2) \Bigg[\int\f{d^dp}{(2\pi)^d}\ d^dz\ d^dz'e^{-i(p_i+k).z'} \  \left( 2  \varepsilon.p \   \right)  \f{e^{ip.(z'-z)}}{p^2+m^2} G[\{\phi(z_j)\}]\Bigg].
\end{align}  
In the above expression, we have used the $\ell$ independent part of both modes and the propagator. We can perform $z'$ integral to obtain $\delta^d(p-p_i-k)$ and use this delta function to do the $p$ integral to get
\begin{align}
  & \sum_{i = 1}^{n}e_i  \left(\f{2\varepsilon.(p_i+k)}{2p_i.k+k^2} \   \right) \int d^dz\ e^{-i(k+p_i).z} \prod_{\substack{j = 1\\ j \neq i}}^{n-1}d^d z_j e^{-ip_j.z_j} (-i)(\nabla_{z_j}^2 - m^2) \Big[  G[\{\phi(z_j)\}]\Big].
\end{align}  
Apart from the leading soft factor, the above expression also contains $O\left(\f{1}{\ell^2}\right)$ pieces coming from $\varepsilon.k$ and $k^2$. Using \eqref{Amodes} and \eqref{gauge}, we obtain the following contribution from the above term to the subleading soft factor
\begin{align}\label{l23}
  &\sum_{i = 1}^{n}e_i\left[-  \f{(d-2)(d-4)}{4\ell^2} \left(\f{2\varepsilon.p_i}{(2p_i.k)^2} \   \right)-\f{(d-4)}{4\ell^2}  \left(\f{\varepsilon_\mu\eta^{\mu\nu}\p_{i\nu}}{p_i.k} \   \right) \right]\Gamma_{n}^{(0)}(\{p_i\}).
 \end{align}
Next, we turn to the $z$ dependent part of \eqref{l21}. It is clear that calculations become much easier if we use the asymmetric form of scalar propagator given in \eqref{fpasym}. In this case, the derivative $\p_\nu'$ appearing in \eqref{l21} will just acts on the exponent in the propagator. Using \eqref{fpasym} in \eqref{l21}, we obtain
\begin{align}\label{l21z}
\int\f{d^dp}{(2\pi)^d}\ d^dz\ d^dz' \left( 2  \varepsilon.p \   \right) \sum_{i = 1}^{n}e_i \Bigg[ & - \frac{d z^2}{4\ell^2} + \f{4m^4}{\ell^2(p^2+m^2)^3}+\f{2m^2(d-3-ip.z)}{\ell^2(p^2+m^2)^2}+\f{(d-2)}{4}\f{z^2}{\ell^2}\nn\\
&+\f{m^2z^2-(d-2)ip.z+(d-2)^2}{2\ell^2(p^2+m^2)}\Bigg]\f{e^{ip.(z'-z)}}{p^2+m^2}e^{-i (p_i + k) z'}\mathcal{V}(\{p_j\})
\end{align} 
where the first term in the square bracket comes from $\sqrt{-g(z)}$ factor. In the above equation, we can perform $z'$ integral to obtain $\delta^d(p-p_i-k)$ and use this delta function to do the $p$ integral. We arrive at
\begin{align}\label{zl21}
\sum_{i = 1}^{n}e_i \int d^dz \left(\f{\varepsilon.p_i}{p_i.k} \   \right)  \Bigg[ & \f{4m^4}{\ell^2(2p_i.k)^3}+\f{2m^2(d-3-i(p_i + k).z)}{\ell^2(2p_i.k)^2}-\f{z^2}{2\ell^2}\nn\\
&+\f{m^2z^2-i(d-2)(p_i +k).z+(d-2)^2}{2\ell^2(2p_i.k)}\Bigg]e^{-ip.z} \ \mathcal{V}(\{p_j\})
\end{align} 
Let us now analyze the $z'$ dependent part of \eqref{l21} which is given by:
\begin{align}\label{gpcon}
\sum_{i = 1}^{n}e_i \int\f{d^dp}{(2\pi)^d}\ d^dz\ d^dz'\ \Bigg[& ( 2 \varepsilon.p) \Bigg(\f{(d-4)}{8}\frac{z'^2}{\ell^2}- \frac{(d-2) z'^2}{4\ell^2} -\f{d(d-1)(d-2)}{24\ell^2m^2}+(d-1)^2\f{ip_i.z'}{8m^2\ell^2}\nn\\
+&(d-1)\f{z'^2}{4\ell^2}+(d^2-1)\f{(p_i.z')^2}{8\ell^2m^2}-\f{d\ ip_i.z'\ z'^2}{4\ell^2}-(3d+1)\f{i(p_i.z')^3}{12m^2\ell^2}\nn\\
-&\f{(p_i.z')^4}{8\ell^2m^2}-\f{(p_i.z')^2 z'^2}{4\ell^2}-\f{m^2 z'^4}{8\ell^2}\Bigg)-\left(\ \frac{\varepsilon.z'}{\ell^2}\right) \Bigg]\f{e^{ip(z'-z)}}{p^2 + m^2}e^{-i(p_i + k) z'}\mathcal{V}(\{p_j\})\ 
\end{align}
where first term is the contribution from gauge field modes $f_k^{*\nu}(z')$ and the second term comes from $\sqrt{-g(z')}$ and $g^{\mu\nu}(z')$ factor. The rest of the terms are $\f{1}{\ell^2}$ contributions to the scalar field mode $g^*_{p_i}(z')$. The last term in the above expression comes from the second piece of \eqref{l21}.

We will now use integration by parts (similar to the one explained in \eqref{intpart}) to remove $z'$ dependence and finally perform $z'$ integral to obtain $\delta^d(p-p_i-k)$ and use this delta function to do the $p$ integral. The lower point amplitude is related to the function $\mathcal{V}(\{p_j\})$ as follows
\be
\Gamma^{(0)}_n=\int d^dz'\ e^{-ip_i z'}\mathcal{V}(\{p_j\}).
\ee

Finally, adding all the contributions from \eqref{l23}, \eqref{zl21} and \eqref{gpcon}, the subleading soft factor turns out to be
\begin{align}
S^{(2)} = \sum_{i = 1}^{n}   e_i &\Bigg(-m^2\frac{(d-4) }{4} \frac{\varepsilon .p_i}{(p_i.k)^3} + i m^2\frac{(d-4) }{4} \frac{\varepsilon .p_i}{(p_i.k)^3}k.\p_{p_i} + \f{m^2}{2} \left(\varepsilon .p_i\frac{k. \p_{p_i}}{ (p_i.k)^3}-  \frac{\varepsilon. \p_{p_i} }{ (p_i.k)^2}  \right)\nn\\
&+ \frac{ (d-4)}{4}\frac{\varepsilon .p_i}{(p_i.k)^2} - \frac{(d-4) }{4} \frac{\varepsilon .p_i}{(p_i.k)^2}p_i .\p_{p_i} + O \left(\f{1}{k}\right)\Bigg)\label{subsoft}
\end{align}
Note that though there are contributions at order $O\left(\f{1}{k^4}\right)$ in \eqref{zl21}, they neatly cancel out once we take into account the corrections due to scalar modes. Consequently, $S^{(2)}$ starts at $O\left( \f{1}{k^3}\right)$.

\subsubsection{Consistency-checks}
\begin{itemize}
\item \textbf{Gauge Invariance}: 
In order to check the gauge invariance of the subleading soft factor $S^{(2)}$ we start with equation \eqref{l21}. The gauge transformation of the polarization vector $\epsilon_\mu$ takes a complicated form due to the fact that gauge field modes are not just plane waves. But the gauge transformation of the mode $f^{*\nu}_k $ is straightforward and is given by 
\be\label{trn}
f^{*\nu}_k \rightarrow f^{*\nu}_k  - i k^\nu.
\ee
We checked that our starting expression, i.e. \eqref{l21} is gauge invariant. It follows that the soft factor $S^{(2)}$ is also gauge invariant. 

\item In $d = 4$, the subleading soft factor simplifies to
\be\label{S24d}
S^{(2)}_{4d} = \sum_{i = 1}^{n}   e_i   \f{ m^2}{2} \left(\varepsilon .p_i\frac{k. \p_{p_i}}{ (p_i.k)^3}-  \frac{\varepsilon. \p_{p_i} }{ (p_i.k)^2}  \right).
\ee
This simplification is related to the fact that in four dimensions,  Maxwell Lagrangian is Weyl invariant and the gauge field modes are just plane waves as seen in \eqref{Amodes}. Under gauge transformation, the polarization vector is shifted by  
\be\label{trn}
\varepsilon^\mu \rightarrow \varepsilon^\mu  - i k^\mu,
\ee
just like the flat space case. It is trivial to check that \eqref{S24d} is invariant under the above transformation.
\item Similar to \S \ref{consist}, we again compare our result with the soft limit of the S-matrix obtained using a new set of scalar modes constructed in Appendix \ref{nmode}. We have checked that the subleading soft factor $S^{(2)}$ for these modes remains the same as in \eqref{subsoft}. This is a strong hint that this soft factor is universal and is tied to an underlying symmetry.
\end{itemize}

\section{Comparison with classical results in $d=4$}\label{clr}
In the previous section, we obtained leading and subleading corrections to the flat space soft photon theorem. In $d=4$, using \eqref{flats},\eqref{S1} and \eqref{subsoft}, the full soft factor is given by 
\begin{align}\label{fullqm}
S_{\text{qu}} = \sum_{i = 1}^{n}   e_i\frac{\varepsilon. {p_i}}{ (p_i.k)} - \f{ m}{\ell} \sum_{i = 1}^{n}   e_i   \  \f{\varepsilon_\mu k_\nu \hat{J}_i^{\mu\nu}}{ (p_i.k)^2} -i \f{ m^2}{2\ell^2} \sum_{i = 1}^{n}   e_i   \  \f{\varepsilon_\mu k_\nu \hat{J}_i^{\mu\nu}}{ (p_i.k)^3} + \cdots \
\end{align}
where $\hat{J}_i^{\mu\nu}$ is the angular momentum operator which takes the form
$$\hat{J}_i^{\mu\nu} = i(p_i^\mu \p_{p_i}^\nu-p_i^\nu \p_{p_i}^\mu).$$

In this section, we will compare the above result with previously known results. In \cite{2108}, the classical subleading soft factor in $4d$ de Sitter spacetime was obtained by calculating electromagnetic radiation emitted in a classical scattering process. The O$\left(\f{1}{\ell^2}\right)$ corrections to the classical soft radiative field $\tilde A_\mu$ emitted in the scattering of $n$ particles with asymptotic momenta $p_i$ was shown to have the following form\footnote{See Eqn (43) of arXiv version 1 of \cite{2108}.}
\begin{align}\label{class}
  \varepsilon.\tilde{A}(\omega,r,\hat{x}) = -\f{1}{4\pi r} \Big[\ \sum_{i = 1}^{n}   e_i\frac{\varepsilon. {p_i}}{ (p_i.k)} - \f{i m^2}{2\ell^2} \sum_{i = 1}^{n}   e_i   \  \f{\varepsilon_\mu k_\nu J_i^{\mu\nu}}{ (p_i.k)^3}\ + O\left(\f{1}{k\ell^2}\right) \Big],  
\end{align}
where $k^\mu=\omega(1,\hat{x})$ and $J_i^{\mu\nu}$ is the classical angular momentum of the $i^{th}$ scattered particle given by $(p_i^\mu x_i^\nu-p_i^\nu x_i^\mu)$. Using \cite{Laddha:2018rle}, we can read off the classical soft factor from \eqref{class} and it is given by
\begin{align}\label{classsoft}
  S_{\text{cl}}=  \ \sum_{i = 1}^{n}   e_i\frac{\varepsilon. {p_i}}{ (p_i.k)} - \f{i m^2}{2\ell^2} \sum_{i = 1}^{n}   e_i   \  \f{\varepsilon_\mu k_\nu J_i^{\mu\nu}}{ (p_i.k)^3}\ + O\left(\f{1}{k\ell^2}\right) .  
\end{align}

Comparing the two results \eqref{fullqm} and \eqref{classsoft} obtained from the S-matrix and classical radiation respectively, one might worry that there is an inconsistency. In particular, the classical soft factor \eqref{classsoft} does not contain any $O\left(\f{1}{\ell}\right)$ corrections while the quantum soft factor\footnote{We call the results obtained from tree-level S-matrix as quantum soft factors.} \eqref{fullqm} receives corrections at this order. We will demonstrate below that this apparent inconsistency goes away by redefining classical momenta $p_i$. In the classical analysis, we can redefine 
\be\label{redf}
p_i \rightarrow c\Big[p_i+\f{am}{\ell} d_i\Big],
\ee
where $c$ is some constant that can be fixed by demanding $p_i^2 = - m^2$. The normalization constant $c$ turns out to be
$$\f{1}{\sqrt{1- 2 a\f{p_i.d_i}{m\ell}-a^2\f{d_i^2}{\ell^2}}}.$$
Under this redefinition of classical momentum, the leading soft factor in \eqref{classsoft} changes to 
\be   e_i\frac{\varepsilon. {p_i}}{ (p_i.k)}\rightarrow e_i\frac{\varepsilon. {p_i}}{ (p_i.k)}+\f{am}{\ell}\Big[e_i\frac{\varepsilon. {d_i}}{ (p_i.k)}-e_i\varepsilon. {p_i}\frac{k. {d_i}}{ (p_i.k)^2}+...\Big]\nn
\ee
i.e.
\be   e_i\frac{\varepsilon. {p_i}}{ (p_i.k)}\rightarrow e_i\frac{\varepsilon. {p_i}}{ (p_i.k)}-\f{am}{\ell}\f{\varepsilon_\mu k_\nu J_i^{\mu\nu}}{ (p_i.k)^2}+ O (k^0)
\ee
For $a =1$, this reproduces the $O\left(\f{1}{\ell}\right)$ piece of quantum soft factor given in \eqref{fullqm}. Hence after this redefinition, $S_{\rm cl}$ becomes equal to $S_{\rm qu}$.

A natural question arises: Is there any allowed redefinition of momentum $p_i$ that can change the subleading soft factor ($S^{(2)}$) as well? Since we can only shift classical momentum $p_i$ by a real function, only 'real' $S^{(1)}$ can be obtained in this way. Hence such a redefinition cannot change the imaginary part of $S^{(2)}$. Moreover, as seen above, these kinds of redefinitions cannot produce $O\left(\f{1}{k^2}\right)$ pieces. This is another hint that $S^{(2)}$ is universal.

\section{Discussion}

In this paper, we studied the perturbative effect of background de Sitter potential on flat space soft photon theorems. We constructed a perturbative S-matrix in a small patch inside the static patch of de Sitter spacetime. As expected, the late time acceleration of scattering particles in the de Sitter background leads to new non-analytic modes in the soft limit of amplitudes as written in \eqref{notation}. 

We first constructed an S-matrix in \eqref{S} for the scattering of orthogonal scalar modes $g_p$ given in \eqref{gp}. Studying the soft limit of scattering amplitudes in minimally coupled scalar QED, we showed that the leading soft correction, i.e. $S^{(1)}$ given in \eqref{S1} is non-universal while we expect that the subleading correction $S^{(2)}$ given in \eqref{subsoft} is universal up to $O(\f{1}{k^2})$. To demonstrate this, we studied soft theorems for the scattering of a different class of scalar modes $f_p$ given in \eqref{fp} and showed that $S^{(2)}$ is same as that for $g_p$ up to $O(\f{1}{k^2})$. Another supporting evidence of this universality is the classical calculation done in \cite{2108}, which agrees with our $S^{(2)}$ in $d=4$. It should be noted that the classical soft radiation does not get any corrections at $O(\f{1}{\ell})$, i.e. $S_{\text{cl}}^{(1)}=0$. This follows from the fact that the effect of background metric starts at $O(\f{1}{\ell^2})$. We recall that $O(\f{1}{\ell})$-correction in the S-matrix calculation arose due to the fact that we had to add $O(\f{1}{\ell})$-terms in the scattering modes $g_p$ to make the modes orthogonal. The classical soft factor and quantum soft factor become equivalent after a redefinition of classical momentum as given in \eqref{redf}. The next step would be to understand the physical meaning of this redefinition. This requires a careful study of the classical limit of the $g_p$ modes, and we leave it for further investigation.

It would be interesting to generalize our analysis to the theories involving non-minimal couplings between matter and the gauge field. Naive power counting suggests that in the de Sitter background, the effects of non-minimal couplings start at $O(k^0)$ in the soft expansion similar to the flat space case.

Since our calculations are perturbative in large $\ell$, they can be easily generalized to Anti-de Sitter space. Hence we can deduce the universal corrections to flat space soft photon factor in AdS by taking $\ell^2 \rightarrow - \ell^2$ in \eqref{l21}. Using \eqref{subsoft}, we obtain
\be
\begin{split}
S^{(2)}_{\text{AdS}} = -\sum_{i = 1}^{n}   e_i &\left(-m^2\frac{(d-4) }{4} \frac{\varepsilon .p_i}{(p_i.k)^3} + i m^2\frac{(d-4) }{4} \frac{\varepsilon .p_i}{(p_i.k)^3}k.\p_{p_i} + \f{m^2}{2} \left(\varepsilon .p_i\frac{k. \p_{p_i}}{ (p_i.k)^3}-  \frac{\varepsilon. \p_{p_i} }{ (p_i.k)^2}  \right)\right.\nn\\
&+ \frac{ (d-4)}{4}\frac{\varepsilon .p_i}{(p_i.k)^2}\left. - \frac{(d-4) }{4} \frac{\varepsilon .p_i}{(p_i.k)^2}p_i .\p_{p_i} + O \left(\f{1}{k}\right)\right)
\end{split}
\ee
Soft photon theorems in AdS have been previously explored in \cite{Banerjee:2022oll,ads}. In their work, they have not included the corrections to the external hard particle states due to AdS potential. We suspect that after considering these effects, their results should agree with ours.

The fact that the form of $S^{(2)}$ is universal raises a natural question: Are these soft modes related to any 'asymptotic' charges? It seems natural from our analysis that the perturbative corrections to flat space asymptotic charges should reproduce the universal parts of the corrected soft factors. These charges ($\mathcal{Q}_\ell$) should be defined on time slices in the static patch. As is well-known, in flat space, the asymptotic charges preserve the boundary conditions near null infinity ($\mathcal{I}^{\pm}$). Similarly, in de Sitter spacetime, one can define horizon charges ($\mathcal{Q}_H$), which preserve the boundary conditions near the horizon of the static patch. An interesting question will be to understand the relationship between the charges $\mathcal{Q}_\ell$ and $\mathcal{Q}_H$. More specifically, does the large $\ell$ limit of $\mathcal{Q}_H$ reduce to $\mathcal{Q}_\ell$? We leave these questions for future works.

 \begin{acknowledgments}
We are extremely grateful to Alok Laddha and Ashoke Sen for numerous discussions and valuable suggestions during the project. We also thank Nabamita Banerjee, Abhijit Gadde, Shiraz Minwalla, Chintan Patel and Trakshu Sharma for discussions. We acknowledge the support of the Department of Atomic Energy, Government of India. SB is thankful for the support of the Infosys Endowment for the study of the Quantum Structure of Spacetime. Finally we would like to thank the people of India for their steady support for research in the basic sciences.

\end{acknowledgments}
\appendix
\section{LSZ for photons}\label{gaugeap}
In this Appendix, we derive the LSZ reduction formula for photons. Using the modes in \eqref{Amodes}, we can expand the gauge field as
\be
A_\mu(x)=\sum_{h=1}^{d-2}\int \f{d^{d-1}k}{(2\pi)^{d-1}}  \ [\ a^h_k f_\mu^{h}\ +\  a^{h\dagger}_k f_\mu^{* h}\ ].\nn
\ee
The above expression can be inverted using the orthogonality of modes using \eqref{inA} to obtain
\begin{align}
a^h_p=(g^{\mu\nu}\ f^{h}_{p\mu},\ A^{h'}_{k\nu})
\end{align}
We will use the above relation to derive LSZ-like formula for creation/ annihilation operators of the gauge field. We will follow the same procedure as done for scalar fields discussed in the main text. Note that
\begin{align}
&a^h_p(T)-a^h_p(-T)=\int^T_{-T} dt\ \p_t a^h_p = (g^{\mu\nu}\ f^{h}_{p\mu},\ A_{k\nu})\nn\\
&=-i\int dt d^{d-1}x \ e^{-\epsilon \f{|\vec{x}|}{R}}\ \p_t \left(\sqrt{-g}\ g^{\mu\nu}(x)\  [\ f^{*h}_{p\mu}(t,\vec{x})\ \p^t A_{k\nu}(t,\vec{x}) -A_{k\mu}(t,\vec{x})\ \p^t f^{*h}_{p\nu}(t,\vec{x})\ ]\right) \nn \\
&=-i\int dt d^{d-1}x\ e^{-\epsilon \f{|\vec{x}|}{R}}\ \p_\rho \left(\sqrt{-g}\ g^{\mu\nu}(x)\  [\ f^{*h}_{p\mu}(t,\vec{x})\ \p^\rho A_{k\nu}(t,\vec{x}) -A_{k\mu}(t,\vec{x})\ \p^\rho f^{*h}_{p\nu}(t,\vec{x})\ ]\right) 
\end{align}
where $\rho$ runs over all spacetime indices. In the second line, we have added the terms at the spatial boundary (which is essentially zero in this case). After taking $\epsilon \rightarrow 0$ limit, the above expression can be evaluated to get
\begin{align}
a^h_p(T)-a^h_p(-T)&=-i\int d^d x \sqrt{-g(x)}\eta^{\mu \nu} f^{*h}_{p\nu} \left(\left(1+ \f{x^2}{\ell^2}\right)\Box \ +\ \f{(4-d)}{2\ell^2}\ [x.\p +1]\right)A_\mu (x). \nn\\ 
\end{align}
For notational brevity, we define
\begin{equation}\label{Dx}
\mathcal{D}_x = \left(1+ \f{x^2}{\ell^2}\right)\Box \ +\ \f{(4-d)}{2\ell^2}\ [x.\p +1],    
\end{equation}
so that we have
\begin{align}\label{lszA}
a^h_p(T)-a^h_p(-T)&=-i\int d^d x \sqrt{-g(x)}\eta^{\mu \nu} f^{*h}_{p\nu} (\mathcal{D}_x) A_\mu (x) 
\end{align}
Using \eqref{photonG}, we find that $\mathcal{D}_x$ acts on Greens function as follows:
\begin{align}\label{G11}
\mathcal{D}_x G_{\mu,\nu} (x,x') &= i\eta_{\mu\nu}\delta^d(x-x')\left[1+\f{d}{4}\f{x^2}{\ell^2}\right] \nn\\
&=  i\eta_{\mu\nu}\f{\delta^d(x-x')}{\sqrt{-g(x)}}
\end{align}

\section{Different class of scalar modes}\label{nmode}
In this Appendix, we construct a different set of solutions of scalar field equation of motion in de Sitter spacetime which behave as plane waves in $\ell \rightarrow \infty$ limit. In the main text, we computed soft factors using the scalar modes $g_p$ given in \eqref{gp}, but those modes are not the only modes with the required flat space limit. 

In $d$ spacetime dimensions, the scalar field equation of motion is given by
\be 
[\nabla^2-m^2]\phi = \left(1+\f{x^2}{2\ell^2}\right)\Box\phi +\f{2-d}{2\ell^2}x.\p\phi-m^2\phi=0.\label{scalareomB}\ee
where we have just kept $O(\f{1}{\ell^2})$ terms in the perturbation theory. The following modes
\begin{align}
f_p=\f{e^{ipx}}{\sqrt{2E_p}}\ \left(1+(d-1)\f{x^2}{4\ell^2}  -\f{ip.x\ x^2}{4\ell^2}-\f{i(x.p)^3}{6m^2\ell^2} + \f{c}{\ell^2}\left((d-1)\f{p.x }{2 m}+\f{imx^2}{2 }+\f{i(p.x)^2}{2 m}\right)\right), \label{fp}    
\end{align}
with
\be
 p^2=-m^2 + \frac{d(d-1) }{2 \ell^2}
\ee
solve \eqref{scalareomB} with any arbitrary constant $c$. Notice that in $\ell \rightarrow \infty$ limit, the above solution behaves as a plane wave with the usual onshell condition. 

\subsection{Difference between $f_p$ and $g_p$}
There are two main differences between the modes $f_p$ listed above and the modes $g_p$ considered in the main text.
\begin{itemize}
    \item  \textbf{Leading correction}: The first distinction is that the leading corrections in the modes $f_p$ start at $O\left(\f{1}{\ell^2}\right)$ whereas the modes $g_p$ receive corrections even at $O\left(\f{1}{\ell}\right)$. Since the scalar field EOM does not receive any corrections at order $\f{1}{\ell}$, the  $O\left(\f{1}{\ell}\right)$ piece of $g_p$ satisfies
    $$\Box \  g^{(1)}_p = 0,$$
    where  $g^{(1)}_p$ is given by 
    $$ g^{(1)}_p = \f{e^{ipx}}{\sqrt{2E_p}}\Big[(d-1)\f{p.x }{2 m}+\f{imx^2}{2 }+\f{i(p.x)^2}{2 m}\Big].$$
    Hence this term can be added to $f_p$ at any order, and the resulting mode will still solve \eqref{scalareomB}. Notice that the constant $c$ in \eqref{fp} also appears with the same term at subleading order. At $O\left(\f{1}{\ell^2}\right)$, the difference between $g_p$ and $f_p$ also satisfies homogeneous equation i.e. $$\Box (g^{(2)}_p - f^{(2)}_p) = 0.$$ 
    \item \textbf{Orthogonality}: The second important difference between the solutions $f_p$ and $g_p$ is orthogonality. Unlike $g_p$, the modes $f_p$ are not orthogonal under the Klein Gordon inner product defined in \eqref{in} i.e.
    \be
    (f_{p}(x),f_p(y)) \neq  \frac{\delta^{d-1}(\vec{x}- \vec{y})}{\sqrt{-g}}
    \ee
    It turns out that the $O\left(\f{1}{\ell}\right)$ piece is necessary to obtain orthogonal modes. This is precisely the reason why we worked with modes $g_p$ to define S-matrix in the main text.
\end{itemize}

\subsection{Soft Photon Theorem with modes $f_p$}
Even though the modes $f_p$ are non-orthogonal, we can define states using these modes as follows :
\begin{align}
\left(a^\dagger_{f_p}(T)-a^\dagger_{f_p}(-T)\right)\ |0\rangle &=i\int^T_{-T} dt d^{d-1}x\ \sqrt{-g}\  f_p^*\ [\nabla^2 - m^2]\ \phi \ |0\rangle .
\end{align}
The corresponding states are not orthogonal. Nevertheless, it is interesting to study the soft limit of the S-matrix for the scattering of these modes. We computed the soft factors for $f_p$ following the calculation in \S \ref{corrections}, and we obtained the following results:
\begin{itemize}
    \item The leading soft factor $S^{(1)}$ defined in \eqref{notation} is zero as these modes do not have any terms at $O\left(\f{1}{\ell}\right)$. This shows that $S^{(1)}$ is not universal.
    \item The subleading soft factor up to $O(\f{1}{k^2})$ turns out to be
    \be
\begin{split}
S^{(2)} = \sum_{i = 1}^{n}   e_i &\left(-m^2\frac{(d-4) }{4} \frac{\varepsilon .p_i}{(p_i.k)^3} + i m^2\frac{(d-4) }{4} \frac{\varepsilon .p_i}{(p_i.k)^3}k.\p_{p_i} + \f{m^2}{2} \left(\varepsilon .p_i\frac{k. \p_{p_i}}{ (p_i.k)^3}-  \frac{\varepsilon. \p_{p_i} }{ (p_i.k)^2}  \right)\right.\nn\\
&+ \frac{ (d-4)}{4}\frac{\varepsilon .p_i}{(p_i.k)^2}\left. - \frac{(d-4) }{4} \frac{\varepsilon .p_i}{(p_i.k)^2}p_i .\p_{p_i} + O \left(\f{1}{k}\right)\right).
\end{split}
\ee
This precisely matches the subleading soft factor computed using the modes $g_p$ in \eqref{subsoft}. It is worth mentioning that $S^{(2)}$ is independent of $c$ appearing in the modes \eqref{fp}. 
\end{itemize}

\bibliographystyle{JHEP}
\bibliography{refer.bib}
\end{document}